\documentclass{article}

\usepackage[english]{babel}
\usepackage[T1]{fontenc}
\usepackage[utf8]{inputenc}
\usepackage{subfig}
\usepackage{graphicx}
\usepackage{xcolor}
\usepackage{float}

\usepackage{tgheros}
\usepackage{varwidth}
\usepackage{amsmath,amssymb,amsthm,textcomp}
\usepackage{enumerate}
\usepackage{multicol}
\usepackage{tikz}
\usepackage{hyperref}
\usepackage{fancyhdr}

\usepackage{geometry}
\usepackage{amsmath}
\usepackage{amsfonts}
\usepackage{color}
\usepackage{amssymb,amsthm}
\usepackage{mathtools}
\usepackage{mathrsfs}
\usepackage{dsfont}
\usepackage{hyperref}    
\usepackage[linesnumbered,vlined]{algorithm2e} 
\usepackage{cleveref}    

\usepackage{listings}
\usepackage{natbib}
    \bibpunct[, ]{(}{)}{,}{a}{}{,}%

\newcommand{\ignore}[1]{}

\author{V. Guigues\\EMAp\\ FGV \and 
A. J. Kleywegt\\Industrial and\\ Systems Engineering\\ Georgia Institute\\ of Technology \and
G. Amorim\\EMAp\\FGV \and
A. Krauss\\Department of\\ Informatics\\ PUC-RJ \and
V. H. Nascimento\\EMAp\\FGV
}

\title{LASPATED: A Library for the Analysis of Spatio-Temporal Discrete Data (User Manual)}

\date{\today}

\begin{document}

\maketitle

\begin{abstract}
This is the User Manual of the LASPATED library.
This library is available on GitHub (at  {\url{https://github.com/vguigues/LASPATED}})) and provides a set of tools to analyze spatio-temporal data.
A video tutorial for this library is available on Youtube.
It is made of a Python package for time and space discretizations and of two packages (one in Matlab and one in C++) implementing the calibration of the probabilistic models for stochastic spatio-temporal data proposed in the companion paper \cite{laspatedpaper}.
\end{abstract}


\section{Introduction}

\paragraph{}
LASPATED is a library
for the analysis of spatio-temporal data. It has a Python  module for space and time discretization and a module for calibration of probabilistic models for such data. LASPATED takes as input a file of events with spatio-temporal information (a .csv file for example) or the corresponding Python 
DataFrame object. Each event has a location, a timestamp, and possibly a set of features.\footnote{We use the terms events and arrivals with the same meaning in this manual.} LASPATED provides three different types of space discretization (in rectangles, hexagons, or customized) and several types of time discretization for such events. 
LASPATED also allows
combining information from two
different space discretizations. It computes the areas of intersection of all the subregions from the two discretizations. Given some attribute (for instance, the population) from the first discretization, LASPATED also computes the values of this attribute in the regions of the second discretization.

Time discretizations provide different partitions of time, for instance
periodic with time windows of equal duration,
periodic with time windows of different durations, or more general partitions, see
Section \ref{sec:timedis}. 

The calibration module considers a discrete spatio-temporal time series, possibly built by the discretization module. The underlying model assumes  the number
of arrivals of type $c$, for zone $i$, and time window $t$, is Poisson with some
intensity $\lambda_{c,i,t}$. 
Two strategies are implemented for the calibration of these intensities: maximizing
a regularized Likelihood with penalizing weights obtained by cross validation and
a model with covariates.
The corresponding optimization problems
are solved using projected gradient with line search along the feasible direction.

The outline of this manual is the following.
Sections~\ref{sec:init} to~\ref{sec:retriev} describe the discretization tools.
In Section \ref{sec:init}, we explain how to load data.
Time and space discretization functions 
are respectively discussed
in Sections \ref{sec:timedis} and \ref{sec:spaced}. 
Additional discretization features are provided in Section
\ref{sec:adddisc} while Section \ref{sec:retriev}
explains how to access discretized data and how to write
such data in files.
Section \ref{sec:modelcalibration} discusses the Matlab and C++ functions for the calibration of the probabilistic models. Our code is tested using GCC 11.4 and Python 3.10. The projection with covariates was solved using Gurobi 10.0.3.

\section{Python installation guide}

LASPATED python library can be installed with pip. After downloading the source code, go to laspated directory and run:

\begin{verbatim}
    $ pip install . 
\end{verbatim}

Throughout this manual, the \$ sign represents the command line interface chosen by the user, whether it is Windows or Linux. The Docker container (see section~\ref{sec:docker}) uses Linux Ubuntu. After running pip, the Python module and all its dependencies will be automatically installed.

\section{Initializing a data aggregator and adding events}\label{sec:init}

LASPATED library is imported with
command 
{\textbf{import laspated}}. 
The library uses a \textit{DataAggregator} object of spatial and temporal
information. 

There are multiple options when using latitude-longitude numbers to represent locations. When initializing the data aggregator object, we should specify the CRS (Coordinates Reference System) used in the data. It is important to note that our data can be passed to different CRS, but LASPATED will use the specified CRS when aggregating information, for instance we can choose
app = spated.DataAggregator(crs="epsg:4326"), see Example \ref{list:0} below.

The observations we are interested in
are spatio-temporal observations which have
a date and time information, a location information,
and possibly additional features.
As an example, consider the following 19 first lines
of a CSV file reporting emergency calls to
Rio de Janeiro emergency health service
(a given line of this file available on the GitHub page of the project corresponds to a call and the information we have on the calls are the time
of the call, the location given by latitude and longitude, and the priority of the call coded by an integer between 0 and 3):
\begin{verbatim}
date_time;priority;long;lat
01/01/2016 02:09;3;-43.33366;-22.86939
01/01/2016 02:13;0;-43.67589;-22.87618
01/01/2016 02:17;1;-43.29333;-22.90662
01/01/2016 02:28;0;-43.54172;-23.01242
01/01/2016 02:44;2;-43.35943;-23.01030
01/01/2016 02:51;0;-43.35715;-22.85183
01/01/2016 02:56;3;-43.63401;-23.00212
01/01/2016 02:57;2;-43.25316;-22.99458
01/01/2016 03:00;0;-43.39274;-22.82499
01/01/2016 03:01;1;-43.36692;-22.86628
01/01/2016 03:03;2;-43.51445;-22.87363
01/01/2016 03:05;2;-43.60198;-22.91049
01/01/2016 03:06;3;-43.17507;-22.96576
01/01/2016 03:07;3;-43.29519;-22.83136
01/01/2016 03:11;3;-43.68198;-22.97946
01/01/2016 03:18;3;-43.18982;-22.96926
01/01/2016 03:24;1;-43.46151;-23.02613
01/01/2016 03:26;3;-43.18726;-22.97490 
\end{verbatim}

This file contains in the first column
a date and time information (column 
{\textbf{date\_time}} with date, hour, and minutes
information), in the second column a priority (an integer between 0 and 3), in the third column  
{\textbf{long}} the longitudes, and in the fourth column
{\textbf{lat}} the latitudes.
To load this data with LASPATED, we first read the corresponding CSV file events.csv with
\begin{verbatim}
events = pandas.read_csv("events.csv").
\end{verbatim}
Then the columns of the file are passed as specific arguments \textbf{datetime\_col}, \textbf{lat\_col} and \textbf{lon\_col}
of LASPATED method {\textbf{add\_events\_data}}.
Additional information can be added inside a list
passed to \textbf{feature\_cols} argument.
In the example above, the additional information
is the priority of the call, provided in column
{\textbf{priority}}. Therefore, the data from
this file is loaded with LASPATED  using

\begin{verbatim}
app.add_events_data(events, datetime_col='date_time', lat_col='lat', 
    lon_col="long", feature_cols=['priority'], 
    datetime_format='%d/%m/%Y %H:%M:%S').
\end{verbatim}

Time-reference data can come in different formats, and the most common ones will be automatically detected and understood by LASPATED.
A specific time column format using the \textbf{datetime\_format} argument can be used, as shown in the example above. This format should be defined based on the datetime library pattern, see \url{https://docs.python.org/3/library/datetime.html#strftime-and-strptime-behavior} for details.

We summarize in Example \ref{list:0} the commands to load the events given in the file events.csv whose 19 first lines are shown above.

\begin{lstlisting}[label={list:0},caption=Building a data aggregator and adding events]
    import laspated as spated
    
    # The data aggregator constructor receives a string with the coordinate reference system
    app = spated.DataAggregator(crs="epsg:4326")

    # Creating a pandas dataframe with events. It must have columns with the timestamp, latitude and longitude and it may have other features
    events = pandas.read_csv("events.csv")

    # The add_events_data method stores the events in the aggregator. The user must provide the names of columns for the timestamps, locations and the list of columns for the features of interest.
    app.add_events_data(events, datetime_col='date_time', lat_col='lat', lon_col="long", feature_cols=['priority'],datetime_format='%d/%m/%Y %H:%M:%S')

    # The data aggregator stores the events added in the events_data dataframe
    # Prints the first 19 observations, corresponding to the 19 observations shown above.
    print(app.events_data.head(19))
\end{lstlisting}

\section{Time discretization}\label{sec:timedis}

\subsection{Periodic with elementary time windows repetition}

LASPATED provides multiple types of time discretization via  \textbf{add\_time\_discretization} method. The simplest discretization
is obtained
using periodic time windows,
where each period 
is divided into
so-called elementary
time windows
of the same duration. 
The duration of this
elementary time window
is specified in the second argument of method
add\_time\_discretization while the time unit of this duration is given in the first argument,
which 
is a string with possible values 
'Y' for years,
'M' for months, 'W' for weeks, 'D' for day, 'H' for hours, 'm' for minutes, and 'S' for seconds. The duration
of a period,  in the time unit of the
first argument of the method, is specified in the third argument of the method.
Therefore, the third argument needs to be
a multiple of the second argument: it is the duration of an elementary time window times the number of elementary time windows in a period.
In Example \ref{list:1},
we consider two such discretizations.
In the first one, we want to count the number of arrivals for every day of the week. Therefore, the first argument of add\_time\_discretization is 'D' (the time unit is a day), the second argument is 1 (an elementary time window 
lasts 1 day), and the third argument is 7 (the one-week period has 7 days).
In the second discretization, we want to count the number of arrivals in time windows
of 30 minutes every day.
Therefore, the first argument of 
add\_time\_discretization is 'm' (the time unit is a minute), the second argument is 30 (an elementary time window 
lasts 30 minutes), and the third argument is 30*48 (the one-day period has 
30*48 minutes).

\begin{lstlisting}[label={list:1},caption=Elementary time windows repetition.]
# 7 days in a week
app.add_time_discretization('D', 1, 7)

# 48 30-minutes slots in a day
app.add_time_discretization('m', 30, 30*48)
\end{lstlisting}

\subsection{Periodic with elementary time windows of different durations}

We can modify the previous time discretization keeping 
periodic observations but
with elementary time windows
of different durations. The same method
add\_time\_discretization will be used with the first argument still providing the
time unit ('Y', 'M', 'W', 'D', 'H', 'm', or 'S').
The second argument now is a list providing the
durations of successive elementary time windows.
For instance, in Example \ref{list:2} which follows, the time
unit is 'M' (month) and the second argument
is the list [3, 4, 2, 1, 2] providing successive elementary
time windows of 3 months, 4 months, 2 months, 1 month,
and 2 months. The third argument is the period, which
should be a multiple of the sum of the durations
specified in the list of the second argument.
In Example \ref{list:2} the period is 12
months meaning that we count the number of observations for January-February-March
in a first group, for
April-May-June-July
in a second group, for 
August-September in a third
group, for October 
in a fourth group, and for
November-December in a fifth group.
In this example, if the last argument was
24 instead of 12 we would have 10 groups
of time observations: observations
for January-February-March
of years 1, 3, 5,.$\ldots$, (with the years counted from the first year of historical data) in a first group, for
April-May-June-July of years 1, 3, 5,.$\ldots$, 
in a second group, for 
August-September 
of years 1, 3, 5,.$\ldots$, 
in a third
group, for October of years 1, 3, 5,.$\ldots$, 
in a fourth group, for
November-December of years 1, 3, 5,.$\ldots$,  in a fifth group,
for January-February-March
of years 2, 4, 6,.$\ldots$, in a sixth group, for
April-May-June-July of years 
2, 4, 6,.$\ldots$,
in a seventh group, for 
August-September 
of years 2, 4, 6,.$\ldots$, 
in a eighth group, for October of years 
2, 4, 6,.$\ldots$,
in a ninth group, and for
November-December of years 
2, 4, 6,.$\ldots$, in a tenth group.

\begin{lstlisting}[label={list:2},caption=Elementary time windows of different durations.]
# this creates an index that repeats yearly (12 months)
# January, February, March => observations indexed by time index 0
# April, May, June, July => observations indexed by time index 1
# August, September => observations indexed by time index 2
# October => observations indexed by time index 3
# November, December => observations indexed by time index 4
app.add_time_discretization('M', [3,4,2,1,2], 12)
\end{lstlisting}

\subsection{Customized time intervals}\label{sec:cust}

We can specify more general time windows like holidays or 
special events defined by the user. LASPATED allows such flexibility by defining a set of customized time indexes for time intervals. The user
needs to provide a DataFrame with the following columns:

\begin{enumerate}
    \item start: index start time. When the period for this index starts. 
    \item end: index end time. When the period for this index ends. 
    \item t: index number. 
    The index of the corresponding 
    time period will receive.
    \item repetition: informs whether the index should be repeated in different years ("yearly").
\end{enumerate}

The customized discretization defined by these parameters will be applied to the data. In Example \ref{list:3}, for data given for
years 2016 and 2017, we define a customized discretization 
that considers additional time windows for holiday January 1 (this
holiday is yearly, and the corresponding observations will receive index 1), and for Carnival.
For Carnival, the observations
will receive index 2 and the 
{\textbf{repetition}} column
is None, since for every year
the specific dates of Carnival are
given (02-06 to 02-11 for 2016 and
02-24 to 03-06 for 2017). Arrivals that do not belong to a customized time interval receive the index zero. Also note that a yearly defined 
customized time interval must begin and end the same year.

\begin{lstlisting}[label={list:3},caption=Customized time index.]
import pandas as pd

time_disc_df = pd.DataFrame([
    ["2016-01-01", "2016-01-01", 1, "yearly"],
    ["2016-02-06", "2016-02-11", 2, None],
    ["2017-02-24", "2017-03-06", 2, None],
], columns=["start", "end", "t", "repetition"])

app.add_time_discretization(time_disc_df)
\end{lstlisting}

\section{Space discretization}\label{sec:spaced}

\subsection{Defining borders}\label{sec:borders}

The first step for space discretization
is to define the border of a region
that contains the arrivals.
The border is the region
that will next be discretized
in space.
This can be done with 
LASPATED either using a 
custom map or a region
automatically built from data
(a rectangular region or an approximate convex hull of arrivals). 

\subsubsection{Custom map}

LASPATED method
add\_max\_borders adds 
a max\_border attribute to object app.
The parameter {\tt{data}} of this method
is a GeoDataFrame object\footnote{see \url{https://geopandas.org/en/stable/docs/reference/api/geopandas.GeoDataFrame.html}}. This GeoDataFrame object must have an attribute 
"geometry" encoding the vertices of the
border and this object is obtained by calling the function
\textbf{gpd.read\_file} which takes as
argument a
Shapefile (with .shp extension) 
of the custom map (region under consideration). 
In Example \ref{list:4} below, we show how to use
add\_max\_borders on two examples, with data
lying
in Rio de Janeiro for the first example and data in New-York for the second.
The respective shape files rj.shp
and ny.shp can be found on the internet and also in 
folders Data/rj and Data/ny of the GitHub project.

\begin{lstlisting}[label={list:4},caption=Adding borders from a custom map.]
import geopandas as gpd
import matplotlib as plt

# GeoDataFrame object read from Rio de janeiro shape file
custom_map = gpd.read_file(r'../Data/rj/rj.shp')
# adding as border
app.add_max_borders(data=custom_map)

# Plotting the border
app.max_borders.plot()
plt.show()

# GeoDataFrame object read from New-York shape file
custom_map = gpd.read_file(r'../Data/ny/ny.shp')
# adding as border
app.add_max_borders(data=custom_map)

# Plotting the border
app.max_borders.plot()
plt.show()
\end{lstlisting}

\subsubsection{Rectangular border and convex hull}

If a customized border is not available, we can use a rectangular border containing all events or an approximate convex hull of the space observations. In the first
case, the argument
of parameter {\tt{method}} of LASPATED method
add\_max\_borders is 
"rectangle" while in the second
case this parameter {\tt{method}} is 
"convex". The following Example \ref{list:5}
shows how to define these borders. Only one of these
two options must be chosen.

\begin{lstlisting}[label={list:5},caption=Automatically defined borders.]
# adding rectangular borders
app.add_max_borders(method="rectangle")

# adding as border the convex hull of observations 
app.add_max_borders(method="convex")

# plotting borders and a sample of the events
app.plot_discretization()
\end{lstlisting}

Once borders are defined, the space discretization step will split the corresponding region
into subregions (depending on the discretization
scheme) and link these regions to the arrivals. 
This step is performed with 
LASPATED method
\textbf{add\_geo\_discretization}.

\subsection{Rectangular discretization}

\begin{figure}
\centering
\begin{tabular}{c}
\includegraphics[scale=0.6]{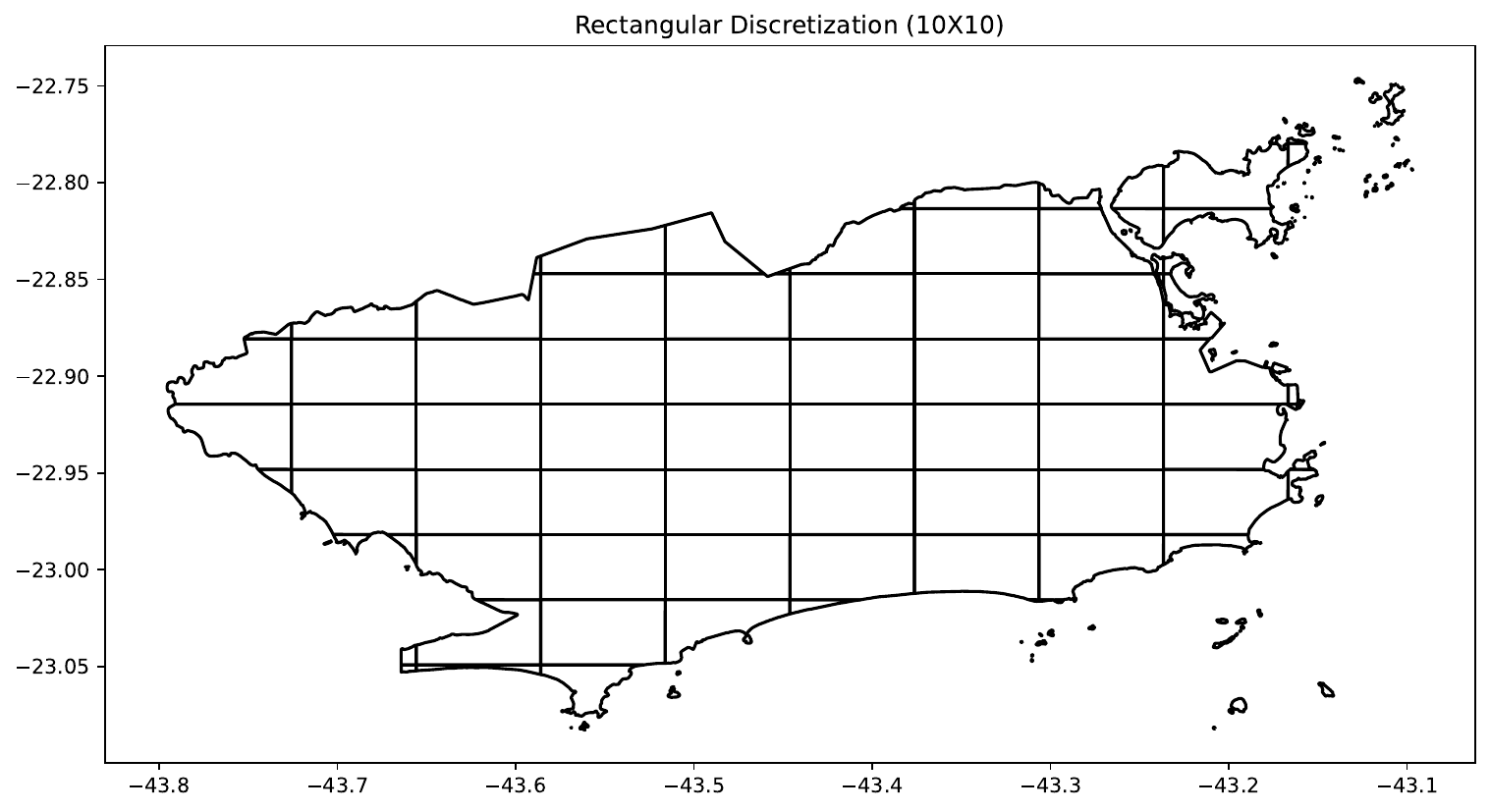}
\end{tabular}
\caption{\label{figurerect1010}
Space discretization of a rectangle containing the city of Rio de Janeiro in $10 \times 10 = 100$ rectangles, 76 of which have nonempty intersection with the city and are represented in the figure.}
\end{figure}

\begin{figure}
\centering
\begin{tabular}{c}
\includegraphics[scale=0.6]{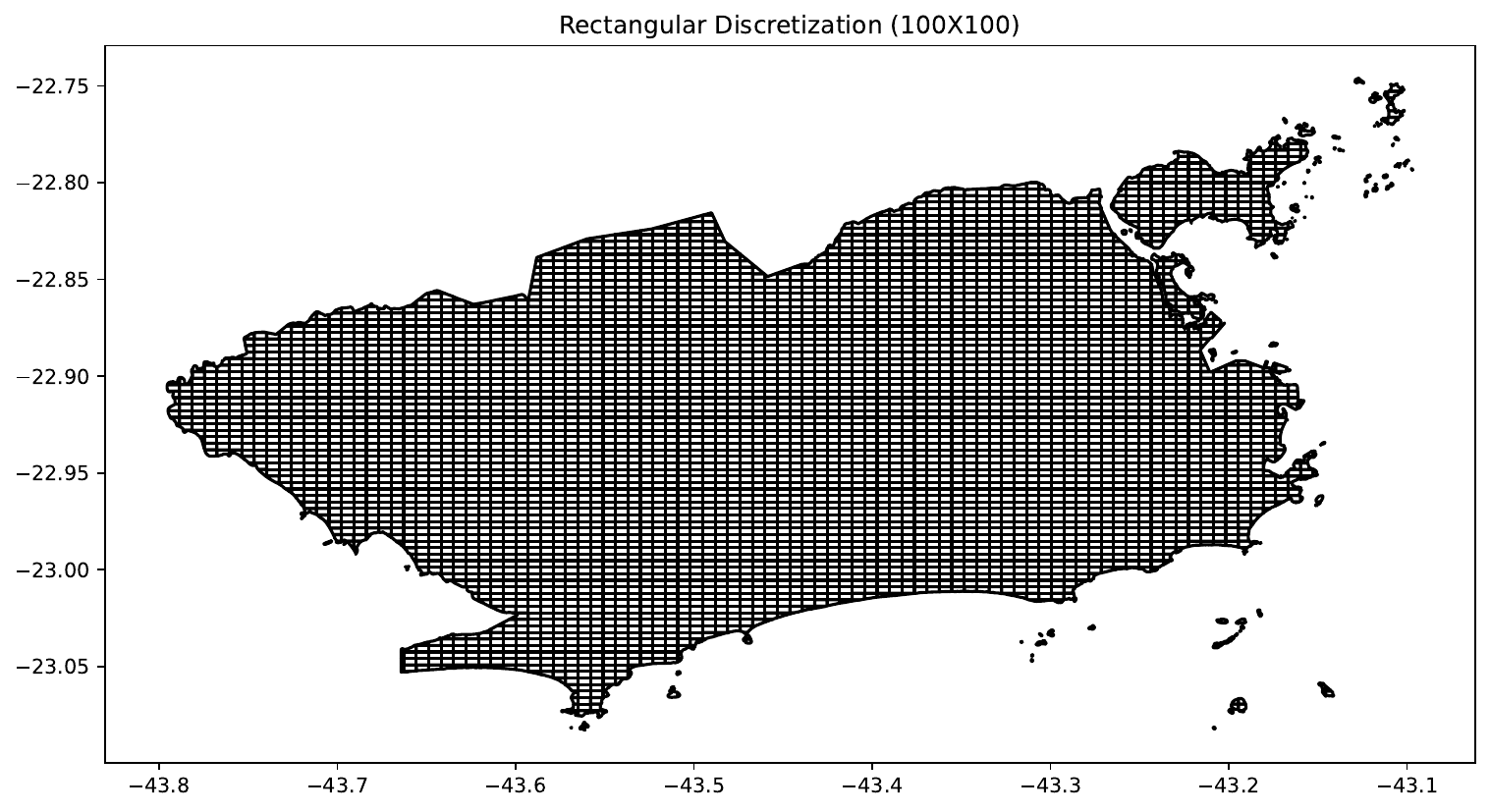}
\end{tabular}
\caption{\label{figurerect100100}
Space discretization of a rectangle containing the city of Rio de Janeiro in $100 \times 100 = 10\,000$ rectangles, 4916 of which have nonempty intersection with the city and are represented in the figure.}
\end{figure}

A rectangular discretization of the studied area defined
in the previous section is obtained
using LASPATED method \textbf{add\_geo\_discretization}
and specifying the value 'R' for parameter
\textbf{discr\_type}, and providing parameters
\textbf{rect\_discr\_param\_x}
and \textbf{rect\_discr\_param\_y} 
which respectively define the 
 horizontal and vertical
discretization steps. 
For instance, to generate
a discretization into
10×30=300 rectangles,
we use the parameters' \textbf{rect\_discr\_param\_x}=10
and \textbf{rect\_discr\_param\_y}=30.  The code of two rectangular
discretizations for the city of
Rio de Janeiro (using the borders of
the city given by the corresponding shape file)
is given in Example \ref{list:6}: the first
discretization is a 10×10 discretization 
in 100 rectangles (resulting
in the rectangular discretization given in
Figure \ref{figurerect1010}) while the second 
discretization is a 100×100 discretization 
in 10,000 rectangles (resulting
in the rectangular discretization given in
Figure \ref{figurerect100100}).

\begin{lstlisting}[label={list:6},caption=Rectangular discretization.]
# Discretization in 10x10=100 rectangles
# The corresponding discretization for Rio de Janeiro city is given in  Figure 1
app.add_geo_discretization(
    discr_type='R',
    rect_discr_param_x=10,
    rect_discr_param_y=10
)
# Plotting the regions
import matplotlib as plt
app.geo_discretization.boundary.plot()
plt.show()

# Discretization in 100x100=10 000 rectangles
# The corresponding discretization for Rio de Janeiro city is given in  Figure 2
app.add_geo_discretization(
    discr_type='R',
    rect_discr_param_x=100,
    rect_discr_param_y=100
)

# Plotting the regions
app.geo_discretization.boundary.plot()
plt.show()
\end{lstlisting}

\subsection{Hexagonal discretization}

LASPATED also provides a hexagonal discretization of the studied area (we have used Uber Python package \textit{H3} to generate the discretization). 

\begin{figure}
\centering
\begin{tabular}{c}
\includegraphics[scale=0.6]{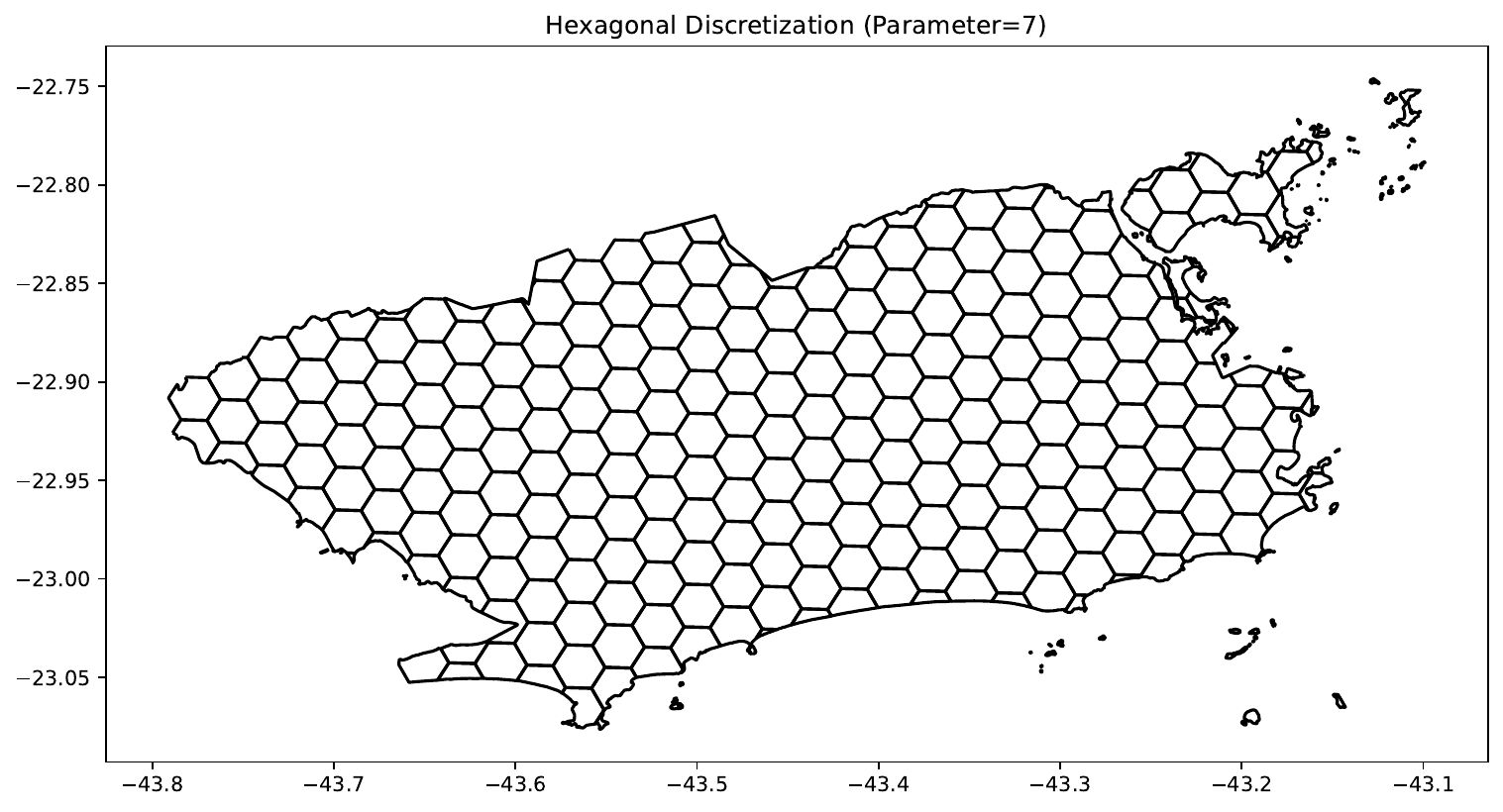}
\end{tabular}
\caption{\label{uber7}Space discretization of a rectangle
containing the city
of Rio de Janeiro in
hexagons using Uber library H3
for space discretization in
hexagons with scale parameter
equal to $7$.}
\end{figure}

This discretization is again obtained with LASPATED method {\textbf{add\_geo\_discretization}} now specifying 'H' for parameter \textbf{discr\_type} and an additional integer parameter textbf{hex\_discr\_param} taking values between 1 and 16, where the smaller the integer the
coarser the discretization.
Figures \ref{uber7} and \ref{uber8} show two hexagonal discretizations for the city of Rio de Janeiro obtained with Example \ref{list:7} below: the first discretization uses scale parameter \textbf{hex\_discr\_param} equal to  7 while the second sets its scale to 8. 

\begin{lstlisting}[label={list:7},caption=Hexagonal discretization.]

# Hexagonal discretization with scale parameter 7
# The corresponding discretization for Rio de Janeiro city is given in  Figure 3
app.add_geo_discretization(
    discr_type='H',
    hex_discr_param=7
)

import matplotlib as plt
app.geo_discretization.boundary.plot()
plt.show()

# Hexagonal discretization with scale parameter 8
# The corresponding discretization for Rio de Janeiro city is given in  Figure 4
app.add_geo_discretization(
    discr_type='H',
    hex_discr_param=8
)

# Plotting the regions
app.geo_discretization.boundary.plot()
plt.show()
\end{lstlisting}

\begin{figure}
\centering
\begin{tabular}{c}
\includegraphics[scale=0.6]{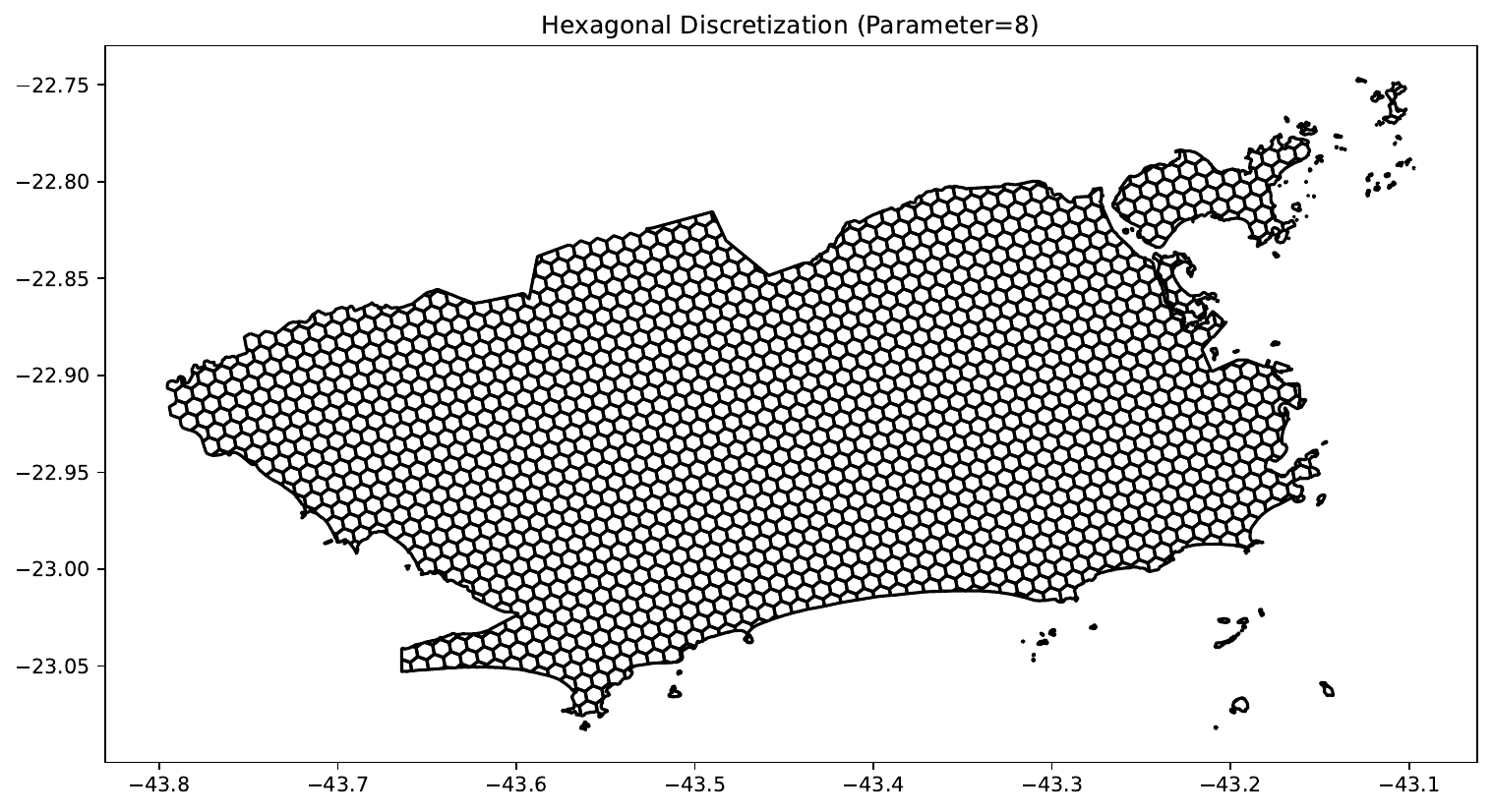}
\end{tabular}
\caption{\label{uber8}Space discretization of a rectangle
containing the city
of Rio de Janeiro in
hexagons using Uber library H3
for space discretization in
hexagons with scale parameter
equal to $8$.}
\end{figure}

\subsection{Customized discretization}

\begin{figure}
\centering
\begin{tabular}{c}
\includegraphics[scale=0.6]{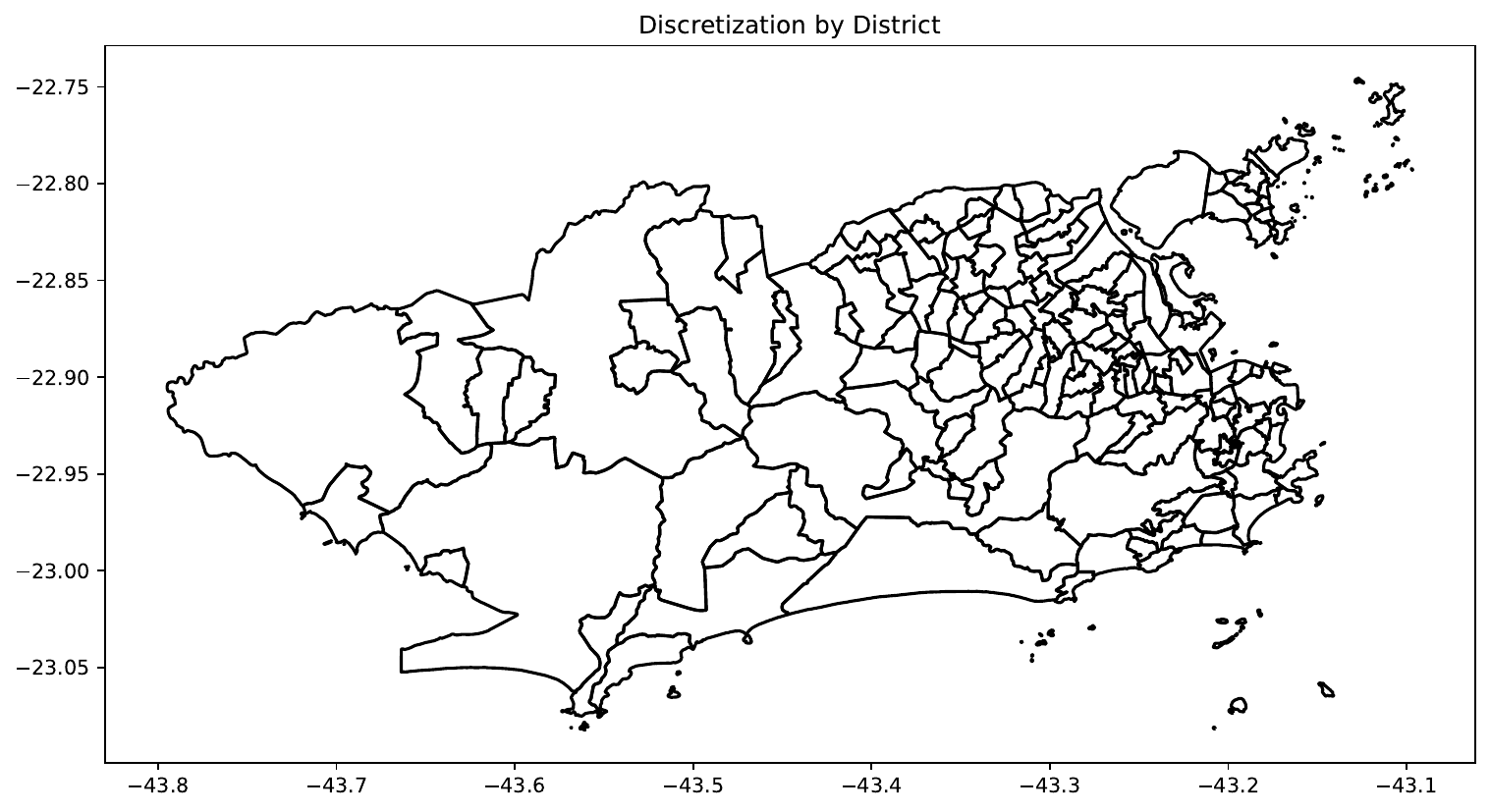}
\end{tabular}
\caption{\label{figuredistrict}Space discretization of the city of
Rio de Janeiro into 160
administrative
districts.\label{distfig}}
\end{figure}

LASPATED also allows discretization with 
customized subregions. 
First, a GeoDataFrame object containing the geometries of each subregion needs to be provided. 
Then the space discretization is obtained
again with method 
{\textbf{add\_geo\_discretization}}
specifying 'C' as \textbf{discr\_type} 
parameter 
and the GeoDataFrame object of the subregions as the additional parameter \textbf{custom\_data}. 
For instance, in Example \ref{list:8} where the studied
area is the city of Rio de Janeiro, 
we take as subregions the administrative
districts.
The corresponding GeoDataFrame object, denoted
by custom\_map in the example, is obtained from the Shapefile 
rio\_neighborhoods.shp of the districts
of the city (this shape file can be found
on the GitHub page of the project).
Figure \ref{figuredistrict} displays the
corresponding customized discretization
by administrative districts.

\begin{lstlisting}[label={list:8},caption=Custom discretization.]
import geopandas as gpd
custom_map = gpd.read_file(r'../Data/rio_de_janeiro_neighborhoods/rio_neighborhoods.shp')
app.add_geo_discretization('C', custom_data=custom_map)

# Plotting the regions
app.geo_discretization.boundary.plot()
plt.show()
\end{lstlisting}

\subsection{Voronoi discretization}\label{sec:voronoi}

Given a set of locations \(L = \{\ell_1, \ell_2,\ldots,\ell_n\}\), LASPATED allows a discretization based on the Voronoi diagram given by the vertices in \(L\). In our context, a Voronoi diagram is a partition of the border, given by the distances to each location in \(L\). In the diagram, zone \(r_i\) contains all locations that are the closest to location \(\ell_i\) than any other location in \(L\). An application of Voronoi diagrams is to define possible covering regions for facilities (hospitals, schools, etc.).

\begin{figure}
\centering
\begin{tabular}{c}
\includegraphics[scale=0.5]{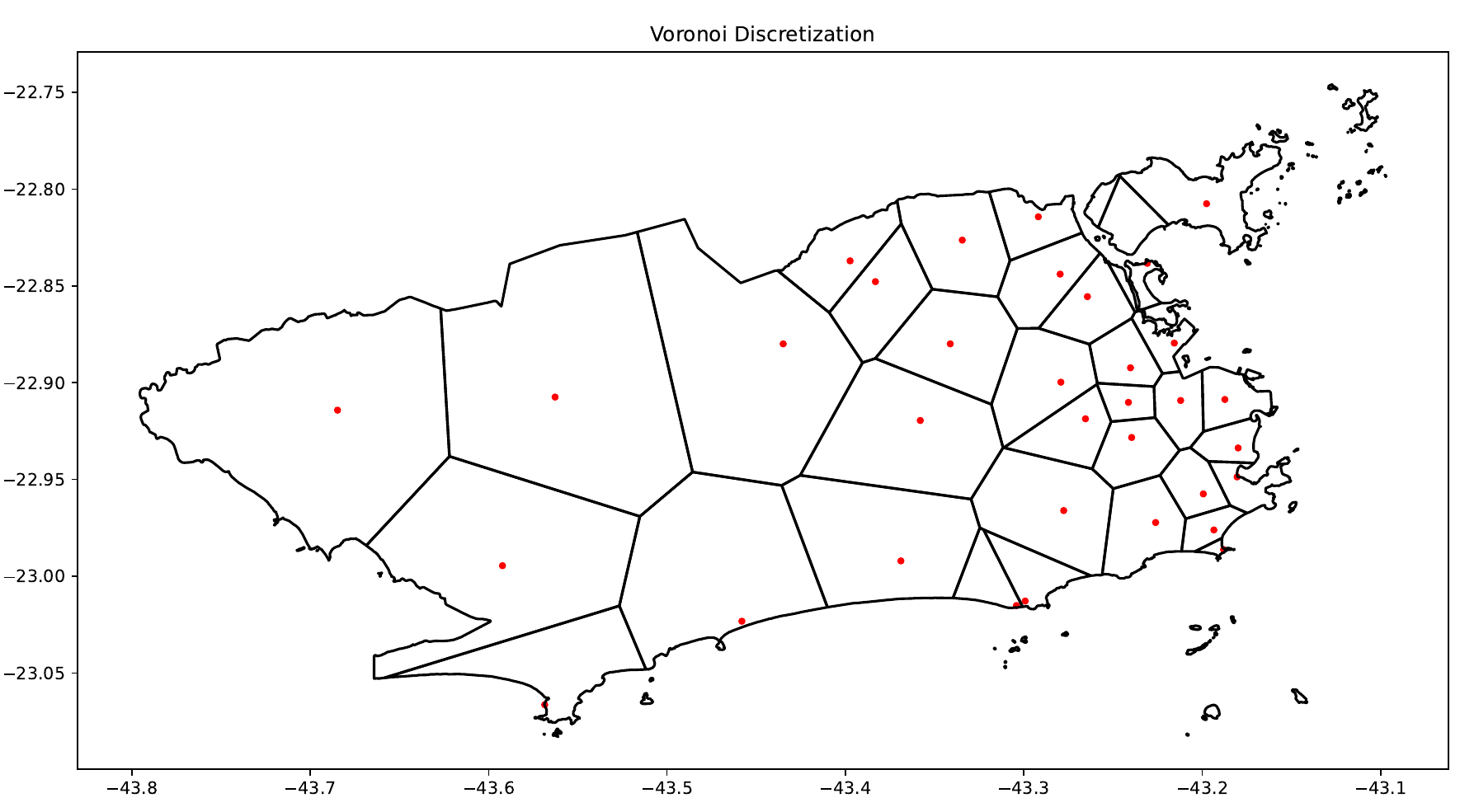}
\end{tabular}
\caption{\label{figurevoronoi} Discretization 
into 34 regions of the city of Rio de Janeiro
given by the Voronoi diagram of ambulance bases in Rio de Janeiro.}
\end{figure}

Example~\ref{list:8.5} loads a GeoDataFrame object corresponding to the ambulance bases in Rio de Janeiro (we read the corresponding Shapefile
bases.shp). Then it obtains the Voronoi discretization using again
Laspated method
add\_geo\_discretization 
by setting 'V' as \textbf{discr\_type} parameter and providing the bases 
as a GeoDataFrame object in parameter \textbf{custom\_data}. Figure~\ref{figurevoronoi} displays the corresponding discretization given by the Voronoi diagram, with the dots corresponding to ambulance base locations.

\begin{lstlisting}[label={list:8.5},caption=Voronoi discretization.]
import geopandas as gpd
bases = gpd.read_file("bases/bases.shp")
app.add_geo_discretization('V', custom_data=bases)

# Plotting the regions
app.geo_discretization.boundary.plot()
plt.show()
\end{lstlisting}

\section{Additional discretization features}\label{sec:adddisc}

In the previous section, we have seen how to define a spatial discretization, say discretization $D_1$.
Let $\mathcal{I}$ be the index
of the regions of this spatial discretization.
Assume we have another discretization
of the studied area, say discretization $D_2$, with 
regions indexed by
$\mathcal{J}$.
In this context, LASPATED provides
the following two functionalities:
\begin{itemize}
\item \textbf{get\_intersection(gpd1, gpd2)}: it computes the areas $\mathcal{A}(i,j)$ of the intersections
of regions 
$i \in \mathcal{I}$ and
$j \in \mathcal{J}$. Parameters gpd1 and gpd2 are the GeoDataFrames corresponding to discretizations \(D_1\) and \(D_2\). The method returns the areas $\mathcal{A}(i,j)$ as a 2-D NumPy array, with the areas measured in square kilometers.
\item \textbf{app.add\_geo\_variable(gpd2,type\_geo\_variable).} Consider a given attribute, for instance
the population and
assume  that for every
region $j \in \mathcal{J}$ of discretization
$D_2$, we have a density $d_j$
of this attribute; the attribute being uniformly distributed
in the region. Let $P_j=d_j \mathcal{A}(j)$ be the value
of the attribute in region $j \in \mathcal{J}$
where $\mathcal{A}(j)$ is the area of region $j$.
When \textbf{type\_geo\_variable} is equal to “feature”, LASPATED computes the value of the
attribute 
$\displaystyle \sum_{j \in \mathcal{J}} d_j \mathcal{A}(i,j)=\displaystyle \sum_{j \in \mathcal{J}} \frac{P_j}{\mathcal{A}(j)} \mathcal{A}(i,j)$ for all regions $i \in \mathcal{I}$. This calculation 
is simply obtained
knowing attributes
$(P_j)$ and 
once areas $\mathcal{A}(j)$
and $\mathcal{A}(i,j)$
(using get\_intersection) have been computed.
The corresponding computation
is done with LASPATED
using method
add\_geo\_variable.
In Example \ref{list:9}, we consider the attribute population.
We first read the corresponding shape file population.shp
generating a GeoDataFrame object
and filter only the columns "population" and "geometry".
This file can be found on the project GitHub page and gives
the population in all the districts of the city of Rio
de Janeiro. Therefore, for this example, discretization
$D_2$ is given by the districts of Rio de Janeiro (see Figure \ref{distfig}). The GeoDataFrame object
filtered as explained above is passed as an argument
of method add\_geo\_variable.
The object app will contain information
about the population of every elementary
cell of discretization $D_1$ (see also the next section).  

If we assume the index set
$\mathcal{J}$ for the second discretization $D_2$
is partitioned into
$k$ index sets $\mathcal{J}_1$, $\ldots$,$\mathcal{J}_k$: $\mathcal{J}=\sum_{i=1}^k \mathcal{J}_i$,
then LASPATED method add\_geo\_variable can be invoked with \textbf{type\_geo\_variable} equal to "area"
can also be used to compute 
$\mathcal{A}_{i \ell}=\displaystyle \sum_{j \in \mathcal{J}_{\ell}} \mathcal{A}(i,j)$
for every $\ell=1,2,\ldots,k$, and $i \in \mathcal{I}$. In Listing~\ref{list:9}, we read a shape file
soil\_use.shp (available on the GitHub page of the project)
providing the different land types (we have 16 of them in
this data) for the city of Rio de Janeiro.
Each elementary cell of this discretization contains
one land type only. We then filter the first two
land types. It then suffices to pass the corresponding
GeoDataFrame object to method add\_geo\_variable.
The object app will contain information
about the areas of land types 1 and 2 for every elementary
cell of discretization $D_1$ (see the next section). 
\end{itemize}


\begin{lstlisting}[label={list:9},caption=Additional discretization features.]
import geopandas as gpd
population = gpd.read_file(r'../Data/regressors/population/population.shp')
population = population[["population", "geometry"]].copy()
app.add_geo_variable(population,type_geo_variable='feature')

land_use = gpd.read_file(r'../Data/regressors/soil_use/soil_use.shp')
land_use = land_use[["type_1", "type_2", "geometry"]].copy()
app.add_geo_variable(land_use,type_geo_variable='area')
\end{lstlisting}

\section{Retrieving and writing discretization data in files}\label{sec:retriev}

LASPATED spatio-temporal 
discretizations
are stored in the dataframe \textbf{app.events\_data} and in the GeoDataFrame \textbf{app.geo\_discretization}.
Dataframe \textbf{app.events\_data} has the same columns as the events dataframe loaded in Section 2, plus time and spatial discretization columns. For each event, the time index discretizations are added with the columns named \textbf{tdiscr\_i}, \($i$ = 0,$\ldots$,T-1\), with T being the number of time index  discretizations defined. Custom column names may be provided in the \textbf{app.add\_time\_discretization} method, with the
default
\textbf{column\_name} parameter. Similarly, the spatial discretization is added with the default name \textbf{gdiscr}. For each event, \textbf{gdiscr} is the index of the region where the event took place. The regions defined in the spatial discretization are fully detailed in the \textbf{app.geo\_discretization} dataframe. For each region, the dataframe stores its index, the list of indexes of neighboring regions, the latitude/longitude of the centroid of the region, and columns with its additional discretization features (for instance population, areas of different land types, see Section
\ref{sec:adddisc} and Example \ref{list:9} for these features). 

LASPATED  method \textbf{get\_events\_aggregated()} returns a NumPy multidimensional array of lists A, where $A[t_0,t_1,\ldots,t_{T-1},g,f_0,f_1,\ldots,f_{F-1}]\) is a list containing the number of events observed in app.events\_data for time discretizations 
$t_0,\ldots,t_{T-1}$, spatial discretization $g$ and features $f_0,\ldots,f_{F-1}$ with $T$ being the number of time discretizations and 
$F$ the number of
features. Example \ref{list:10} shows how to call the method.

\begin{lstlisting}[label={list:10},caption=Aggregating events by discretization.]
app.add_events_data(events, datetime_col='date_time', lat_col='lat', lon_col="long", feature_cols=['priority'])

app.add_time_discretization('m', 30, 30*48) # tdiscr_0
app.add_time_discretization('D', 1, 7) # tdiscr_1

app.add_geo_discretization(
    discr_type='R',
    rect_discr_param_x=10,
    rect_discr_param_y=10
)

events_agg = app.get_events_aggregated()

# Axis are respectively: tdiscr_0, tdiscr_1, gdiscr, priority
print(events_agg.shape)

\end{lstlisting}

The method \textbf{write\_arrivals(path\_arrivals)} calls the \textbf{get\_events\_aggregated()} method and saves A in the path given by \textbf{path\_arrivals}. The method   \textbf{write\_regions(path\_regions)} saves information about the space discretization in the path given by \textbf{path\_regions}, see Example \ref{list:101} and the
corresponding files on the github page of the project.

\begin{lstlisting}[label={list:101},caption=Writing the discretized data to a file.]
app.write_arrivals("arrivals.dat")
app.write_regions("neighbours.dat")
\end{lstlisting}

\section{Model calibration}\label{sec:modelcalibration}

\subsection{Matlab LASPATED calibration functions}

\subsubsection{Calibration of the models for fixed penalties}

LASPATED function laspated calibrates the models
described in Section 2 of the companion paper
\cite{laspatedpaper}. For this function, fixed penalties $w_{i,j}$ and $W_G$ are used and the optimization problems to be solved to calibrate
the models are solved with
the projected gradient method with 
Armijo line search along the feasible
direction.
In this section, we focus on laspated function.

A typical call to laspated is
\begin{center}
\begin{tabular}{l}
[lambda,obj]=laspated(model, nbObservations, nbArrivals, T, R, C, durations, Groups,\\
\hspace*{3.6cm}whichgroup, sigma, iterMax, epsilon, lambda0, params).
\end{tabular}
\end{center}

This function has a variable {\tt{model}}
which is a string 
which is {\tt{noreg}} to calibrate the model without regression and 
{\tt{reg}} to calibrate the model with regression (other values of this string
will generate the message {\tt{Invalid model name}}).

Function laspated has also the following input variables:
\begin{itemize}
\item nbObservations, which gives the number of observations
for every possible combinations of arrival types, time steps,
and regions. More precisely, if {\tt{model}}={\tt{noreg}} then
nbObservations is a 3 dimensional array with nbObservations(c,i,t) the number
of observations for 
arrival type with index c, region with index i, and
time window with index t
(this is parameter $N_{c,i,t}$ of the model without regressors
of \cite{laspatedpaper}, Section 2.)
and if {\tt{model}}={\tt{reg}} then nbObservations(c,d,t,i) is the number
of observations for arrival type $c$, day or holiday $d$,
time window with index t, and region with index i (this is parameter
$N_{c,d,t,i}$ of the model with regressors
of \cite{laspatedpaper}, Section 2).
\item nbArrivals: array providing the number of arrivals.
More precisely,  if {\tt{model}}={\tt{noreg}} then
this is a 3 dimensional array with nbArrivals(c,i,t)
the total number of arrivals observed for time window t,
region i, arrival type c (this is parameter $M_{c,i,t}$ of the model without regressors
of \cite{laspatedpaper}, Section 2)
and if {\tt{model}}={\tt{reg}} then 
nbArrivals(c,d,t,i) is the number
of arrivals for arrival type $c$, day or holiday $d$,
time window with index t, and region with index i
(this is parameter $M_{c,d,t,i}$  of the model with regressors
of \cite{laspatedpaper}, Section 2);
\item T: number of discretized time intervals (this is $|\mathcal{T}|$);
\item R: number of regions (this is $|\mathcal{I}|$);
\item C: number of arrival types (this is $|\mathcal{C}|$);
\item durations: 1-dimensional array with durations(t) the duration
of time interval with index t;
\item Groups: when {\tt{model}}={\tt{noreg}} cell array with Groups\{1,i\} an array
where Groups\{1,i\}(j) is the index of $j$th time interval 
which is in time group with index i. Time groups
could be added in a future version of the model
with regressors but for now set Groups=[] when
{\tt{model}}={\tt{reg}}.
\item whichgroup: 1-dimensional array of size $T$ where whichgroup(t) is the index
of the time group which contains time interval with index t.
Time groups
could be added in a future version of the model
with regressors but for now set whichgroup=[] when
{\tt{model}}={\tt{reg}};
\item sigma: this is the value of parameter $\sigma$
in the pseudo-code for projected gradient with Armijo line search along the feasible
direction given in Section 2 of the companion paper \cite{laspatedpaper};
\item iterMax: maximal number of iterations for projected gradient; this is parameter MaxNumberIteration for the pseudo-code in Section 2 of the companion paper \cite{laspatedpaper};
\item epsilon: parameter $\varepsilon$ for the models ({\tt{reg}} and {\tt{noreg}}) in Section 2 of the companion paper \cite{laspatedpaper};
\item lambda0: initial value $\lambda_0$ of $\lambda$; when
 {\tt{model}}={\tt{noreg}} , this is a 3 dimensional array with lambda0(c,i,t) an
initial guess for the vector of intensities 
$\lambda_{c,i,t}$ for 
 time window with index t, region with index i, arrival type $c$ and lambda0(c,i,t) greater than $\varepsilon$. When {\tt{model}}={\tt{reg}} 
 this is an initial value for $\beta=(\beta_{c,d,t})_{c,d,t}$
which needs to be non negative.
\end{itemize}
Function laspated has also a structure params with parameters
that depend on the model. 

When {\tt{model}}={\tt{noreg}}, then params
has the following fields:
\begin{itemize}
\item params.neighbors: cell array with 
params.neighbors\{1,i\} the list of the indexes of the neighbors
of region with index i;
\item params.weight: 1-dimensional array with 
params.weight(i) the weight $W_G$  corresponding to time group $G$ with index i in the loss function; see
Section 2 of the companion paper \cite{laspatedpaper} for the
expression of the loss function;
\if{
\item params.type(i) is as positive integer coding the type of region with index
i; it is possible not to use
this type information defining all types
equal (say to 1);
}\fi
\item params.distance: 2-dimensional array with 
params.distance(i,j) the distance between the centroids
of regions with indexes i and j;
\item params.alpha: this is the common value of
weights $w_{i j}$ in the loss function (given in 
Section 2 of \cite{laspatedpaper}) which are equal to params.alpha only
when regions with indexes $i$ and $j$ are 
neighbors;
\item params.delta: parameter $\delta$ in stopping criterion (15)
of LASPATED paper \cite{laspatedpaper};
\item params.uppx (resp. params.lowx) is an upper bound (resp. a lower bound)
on $\lambda$.
\end{itemize}
When {\tt{model}}={\tt{reg}}, then params
has the following fields:
\begin{itemize}
\item params.D: number of days plus holidays;
\item params.indexBeta: 
a vector of indexes for $\beta$ such that
$\beta_{c,d,t}(j)$ is given by
\begin{verbatim}
lambda(params.indexBeta(c,d,t,j)) 
\end{verbatim}
(lambda being the optimization variable);
\item params.regressor: params.regressor(j,i) is the value of  regressor
j for region with index i;
\item params.nbRegressors: number of regressors;
\item params.sizex: number of entries in vector $\beta=(\beta_{c,d,t})_{c,d,t}$.
\end{itemize}

The outputs are
\begin{itemize}
\item lambda: when variable {\tt{model}}
is {\tt{noreg}} lambda is an estimated optimal solution. More precisely, this is
a 3 dimensional array with lambda(c,i,t) the optimal  intensity 
$\lambda_{c,i,t}$ for 
 time window t, region i, arrival type $c$. When variable 
 {\tt{model}}
is {\tt{reg}}, lambda
is a 1 dimensional vector with lambda(params.indexBeta(c,d,t,j))
an optimal value for $\beta_{c,d,t}(j)$. 
\item obj: this is the estimated optimal value
along iterations of the optimization method used to solve the optimization problems.
\end{itemize}

\subsubsection{Calibration of penalties}

The calibration of parameters $w_{i,j}$ and
$W_G$ for the loss function
of the model without regressors from Section 2 in \cite{laspatedpaper})
can be done calling function
crossValidation
of LASPATED library. This function uses cross validation
to estimate a value $\alpha=w_{i,j}=W_G$
common to all pairs of neighboring regions $i,j$
and  all groups $G$.
A call to this function is

\begin{verbatim}
    [cputime,weight,lambda]=crossValidation(model, sample, T, R, C, sigma, 
        lambda0, iterMax, proportion, epsilon, durations, Groups, 
        whichgroup, weights, params).
\end{verbatim}

The input variables which have the names of the input variables
of laspated function have the same meaning as before.
Additionally we have the following input variables:
\begin{itemize}
\item sample which is the sample of arrivals.
More precisely, sample is a 
4-dimensional array with 
sample(t,i,c,j) the $j$th observation
for time window $t$, region $i$, and arrival type
$c$;
\item proportion is a real in the interval $(0,1)$
(typically 0.2) which is the proportion of data
used in every estimation block of cross validation
while the proportion (1-proportion) is the proportion of
data used for validation (in cross validation).
\item weights is an array of candidate weights
for $w_{i,j}$ and $W_G$.
\end{itemize}

params is a structure
with the fields
params.neighbors=neighbors, params.distance=distance,
params.alpha=1, which have the same meaning as before.

\subsubsection{Matlab examples}

\par {\textbf{Simulated data, model without regressors.}} File laspatedex1noreg.m of the library contains function laspatedex1noreg
run with the command
$$
{\tt{[weight,lambda,obj]=laspatedex1noreg()}}
$$
which builds
the data corresponding to the example
of Section 6.1 in \cite{laspatedpaper}
and runs laspated calibration function for the
model without regressors
for a set of penalizations. It also runs cross
validation. The output weight
is the best penalization found by cross validation and
lambda is the vector of intensities.

\par {\textbf{Simulated data, model with regressors.}} File 
laspatedex1reg.m of the library contains function laspatedex1reg
run with the command
$$
{\tt{[weight,lambda,obj]=laspatedex1reg()}}
$$
which  builds the data 
corresponding to the example
of Section 6.2 in \cite{laspatedpaper}
and runs laspated calibration
function for this example.\\

\subsection{C++ LASPATED calibration functions}

LASPATED also provides an implementation in C++ of the calibration functions. All code is contained in the header file laspated.h. The user can run the calibration functions by building a model object and a struct with parameters for the projected gradient and cross validation functions. In the following sections we explain how the C++ classes and methods work and show examples of the main functionalities.

\subsubsection{Compilation and dependencies}

C++ calibration functions use the following libraries:

\begin{itemize}
    \item Gurobi (https://www.gurobi.com), an optimization tooblox used only in the model with covariates;
    \item xtensor and xtl libraries.
\end{itemize}

When compiling the code, the user needs to pass the location of the header files for laspated, xtl, xtensor and gurobi, as well as the location for the Gurobi libraries and their respective flags. xtl and xtensor require the C++14 standard and its flag must also be passed. The example below compiles an executable called laspated from a main.cpp that includes laspated. The example considers that the Gurobi solver will be used.

\begin{verbatim}
    $ g++ -o laspated main.cpp -std=c++14 -I<gurobi_header_dir> -I<laspated_header_dir> 
        -I<xtl_header_dir> -I<xtensor_header_dir> -L<gurobi_libs_dir> 
        -lgurobi_c++ -lgurobi100
\end{verbatim}

If Gurobi cannot be acquired, calibration functions for the model without covariates can still be used by passing the flag \textbf{-DUSE\_GUROBI} equal to zero. The following command can be used to compile:

\begin{verbatim}
    $ g++ -o laspated main.cpp -std=c++14 -DUSE_GUROBI=0 -I<laspated_header_dir> 
        -I<xtl_header_dir> -I<xtensor_header_dir> 
\end{verbatim}

We provide five test models in the source files test\_problems.cpp and test\_problems.h in the Model\_Calibration directory. They were used to assert the quality of the projected gradient algorithm.

\subsubsection{Classes and attributes}

The C++ calibration implements a class \textbf{RegularizedModel} that represents the model without covariates and class \textbf{CovariatesModel} that represents the model with covariates. There is also a \textbf{Param} struct, containing all parameters for the projected gradient and cross validation functions. The \textbf{Param} struct has the following members:

\begin{itemize}
    \item double EPS: tolerance used  in the projection algorithm;
    \item double sigma: parameter used in the line search of the projected
            gradient method;
    \item double accuracy: gap used as stopping criterion for the projected
            gradient method;
    \item int max\_iter: maximum number of iterations for the projected
            gradient method;
    \item double lower\_lambda: lower bound for decision variables in the projected
            gradient method;
    \item double upper\_lambda: upper bound for decision variables in the projected
            gradient method;
    \item double beta\_bar: initial step size for the projected gradient
            method;
    \item double cv\_proportion: proportion of samples used in each sub-iteration of the
            cross validation;
    {\color{blue}
    \item relax\_empirical\_fix: if set to true, relaxes the constraint that makes the sum of estimated rates per class equal to the empirical rate by class.}
\end{itemize}

Class \textbf{RegularizedModel} has the following attributes:

\begin{itemize}
    \item param: reference for some parameter object;
    \item nb\_observations: number of observations for each type-zone-time index;
    \item nb\_arrivals: number of events for each type-zone-time index;
    \item durations: duration, in hours, for each time index;
    \item groups: description of time groups. groups[i][j] is the j-th time index
        in the i-th group;
    \item weights: time regularization weight for each time group;
    \item alpha: space regularization weight matrix of zones;
    \item distance: distance matrix (distance between zones);
    \item type\_region: space groups. type\_region[r] is the type of subregion r;
    \item neighbors: neighborhood description. neighbors[i][j] is the j-th
        neighbor of subregion i;
    \item which\_group: groups of each time index. which\_group[d][t] is the index of the time group corresponding to day d and time interval t within day d;
    \item C: number of event types;
    \item R: number of zones;
    \item T: number of periods.
\end{itemize}

The type of each attribute is defined according to the constructor method:

\begin{verbatim}
    RegularizedModel(xt::xarray<int> &nb_observations, 
                    xt::xarray<int> &nb_arrivals,
                    std::vector<double> &a_durations,
                    std::vector<std::vector<int>> &a_groups,
                    std::vector<double> &a_weights, xt::xarray<double> &a_alphas,
                    xt::xarray<double> &a_distance, std::vector<int> &a_type,
                    std::vector<std::vector<int>> &a_neighbors, Param &a_param).
\end{verbatim}

The \textbf{CovariatesModel} has the following attributes:

\begin{itemize}
    \item env: Gurobi environment object;
    \item param: reference for some parameter object;
    \item nb\_observations: number of observations for each type-zone-time index;
    \item nb\_arrivals: number of events for each type-zone-time index;
    {\color{blue}
    \item durations: duration, in hours, for each time index \((d,t)\).}
    \item regressors: array of regressor values. regressor(j,r) is the value of the j-th regressor for
          zone r;
    \item C: number of event types;
    \item D: number of days;
    \item T: number of periods;
    \item R: number of zones;
    \item nb\_regressors: number of regressors;
\end{itemize}

with corresponding constructor:

\begin{verbatim}
    CovariatesModel(xt::xarray<int> &nb_observations, 
                xt::xarray<int> &nb_arrivals, 
                xt::xarray<double>& a_durations, xt::xarray<double> &reg, 
                Param &param).
\end{verbatim}

Both model classes have the same methods, with different implementations. The methods are used in the calibration functions and to assess the quality of its solutions:

\begin{itemize}
    \item double f(xt::xarray<double> \&x): returns the value of the objective function at x,
\item xt::xarray<double> gradient(xt::xarray<double> \&x): returns the gradient of f at x,
    \item xt::xarray<double> projection(xt::xarray<double> \&x): returns the projection of x onto the set defined by the constraints of the model with covariates,
    \item bool is\_feasible(xt::xarray<double> \&x): returns true if x is in the feasible set defined by the constraints of the model with covariates and false otherwise,
    \item double get\_rhs(xt::xarray<double> \&grad, xt::xarray<double> \&dir): returns the value of the directional derivative given by gradient grad and direction dir.
    \item double get\_lower\_bound(xt::xarray<double> \&x, xt::xarray<double> \&grad): returns objective function lower bound given x and gradient grad.
    \item double average\_rate\_difference(xt::xarray<double> \&x1,xt::xarray<double> \&x2): returns average absolute difference in rates between x1 and x2.
\end{itemize}

\subsubsection{Calibration of the model without regressors}

Given the objects described in the previous section, the function:

\begin{verbatim}
    template <typename Model> xt::xarray<double> 
        projected_gradient_armijo_feasible(Model &model, Param &param, 
            xt::xarray<double> &x)
\end{verbatim}

solves the model using the projected gradient method, with parameters given by the param object and with initial solution x. The function uses C++ templates, meaning that Model can be replaced with \textbf{RegularizedModel} or \textbf{CovariatesModel}. The function returns the best solution found in the projected gradient method.

Example~\ref{list:11} calls the calibration method for a model without covariates.
For simplicity, we omitted the code that initializes all the model attributes. We begin including laspated header and using the laspated namespace. Next, we build a \textbf{RegularizedModel} object m, a \textbf{Param} object param and then run the projected gradient method, storing the result in the xarray lambda.

\begin{lstlisting}[label={list:11},caption=Calibration of the model without regressors in C++]

#include "laspated.h"
// All laspated functions in namespace laspated
using namespace laspated;


int main(int argc, char* argv[]){
    // Load variables nb_observations, nb_arrivals, durations, 
    // groups, weights, alphas, distance, type_region, neighbors. 
    // Builds param object with default parameters.
    Param param;
    RegularizedModel m(nb_observations, nb_arrivals, durations, groups, weights, alphas, distance, type_region, neighbors, param);

    // Initial solution
    xt::xarray<double> lambda0 = 0.1*xt::ones<double>(nb_observations.shape());
    // Runs the solver
    xt::xarray<double> lambda = projected_gradient_armijo_feasible<RegularizedModel>(m, param, lambda0);

    return 0;
}
\end{lstlisting}

\subsubsection{Cross Validation}

Cross validation can also be run for the model without regressors.
This function allows us to choose the penalization parameters of the model
without regressors. The 
corresponding cross\_validation function has the following signature:

\begin{verbatim}
    CrossValidationResult cross_validation(Param &param, RegularizedModel &model,
                                       xt::xarray<int> &sample,
                                       std::vector<double> &group_weights)
\end{verbatim}

where CrossValidationResult is the following struct:

\begin{verbatim}
    typedef struct {
        double cpu_time;
        double weight;
        xt::xarray<double> lambda;
    } CrossValidationResult;
\end{verbatim}

Example~\ref{list:13} shows how to execute the cross validation. The user must set the parameter {\tt{cv\_proportion}} \(\in (0,1]\) in the \textbf{Param} object, build a \textbf{RegularizedModel} object and provide a vector of {\tt{cv\_weights}}. The cross validation will evaluate how the model performs considering each weight in {\tt{cv\_weights}} as both the space and time regularization penalties. The returned struct  contains the time spent performing cross validation, the best weight and the solution of the model considering the best weight. Note that the alphas and weights passed when building m are ignored in the cross validation, being replaced with each weight in {\tt{cv\_weights}}.

\begin{lstlisting}[label={list:13},caption=Cross validation in C++]
#include "laspated.h"
#include <vector>

using namespace laspated;
using namespace std;

int main(int argc, char* argv[]){
    // After loading variables nb_observations, nb_arrivals, 
    // durations, groups,weights, alphas, distance, 
    // type_region, neighbors
    // Builds param object with default parameters.
    Param param;
    // Sets proportion for cross validation
    param.cv_proportion = 0.2;
    RegularizedModel m(nb_observations, nb_arrivals, durations, groups, weights, alphas, distance, type_region, neighbors, param);
    vector<double> cv_weights{0.1,1,10};
    CrossValidationResult result = cross_validation(param, m, cv_weights);
	
    return 0;
}

\end{lstlisting}

\subsubsection{Calibration of the model with regressors}

Example~\ref{list:12} performs the calibration of a model with covariates.
Function {\tt{projected\_gradient\_armijo\_feasible}} can still be used for projections. We pass both the class \textbf{CovariatesModel} and an instance of this class to the function. The CovariatesModel object must be built using the observations, arrivals and covariates (regressors) data. The {\tt{projected\_gradient\_armijo\_feasible}} returns the estimated $\beta$ variables.

\begin{lstlisting}[label={list:12},caption=Calibration of the model with regressors in C++]
#include "laspated.h"
// All laspated functions in namespace laspated
using namespace laspated;


int main(int argc, char* argv[]){
    // Load variables nb_observations, nb_arrivals, and regressors. 
    // Builds param object with default parameters.
    Param param;
    //Builds model object
    Covariates m(nb_observations, nb_arrivals, regressors);

    // Initial solution
    xt::xarray<double> beta0 = 0.1*xt::ones<double>(nb_observations.shape());
    // Runs the solver
    xt::xarray<double> beta = projected_gradient_armijo_feasible<CovariatesModel>(m, param, beta0);
    
    return 0;
}
\end{lstlisting}

\subsubsection{Laspated C++ app}
\label{sec:laspated_cpp}

To those unfamiliar with C++, we also provide a laspated binary app that allows running the calibration functions. The user must provide a set of files describing the model and a set of configuration options, either via a configuration file or command line. The options describe the algorithmic parameters, model attributes, and file locations. The valid options are the following:

\begin{itemize}
    \item double EPS: tolerance used in the projection algorithm;
    \item double sigma: parameter used in the line search of the projected
            gradient method;
    \item double accuracy: gap used as stopping criterion for the projected
            gradient method;
    \item int max\_iter: maximum number of iterations for the projected
            gradient method;
    \item double lower\_lambda: lower bound for decision variables in the projected
            gradient method;
    \item double upper\_lambda: upper bound for decision variables in the projected
            gradient method;
    \item double beta\_bar: initial step size for the projected gradient
            method;
    \item double cv\_proportion: proportion of samples used in each iteration of cross validation;
    \item output\_file: path where the laspated app will save the estimated intensities.

    \item string model\_type: the type of model being solved, must be either "reg" (for the model with covariates) or "no\_reg" (for the model without covariates);
    \item string method: the type of method being used. For the model with covariates, only "calibration" is supported (running the projected gradient algorithm once). For the model without covariates, can be either "calibration" or "cross\_validation";
    \item string algorithm: the algorithm used. Can be either "feasible" for Armijo search  along the feasible direction or "boundary" for Armijo search along the boundary. However "cross\_validation" only supports the "feasible" option;
    \item string info\_file: path of file containing general information about the model. See below for a precise description of the corresponding file.
    \item arrivals\_file: path of file containing the samples of arrivals. See below for a precise description of the corresponding file.
    \item neighbors\_file: path of file containing the description of each zone, including its type, its covariates, and its neighbors. See below for a precise description of the corresponding file.
    \item durations\_file: path of file containing the durations for each time index. See below for a precise description of the corresponding file.
    \item alpha\_regions\_file: path of file containing the space penalizations \(R \times R\) matrix. See below for a precise description of the corresponding file.
    \item time\_groups\_file: path of a file containing a description of time groups and the corresponding penalties in the regularized model. See below for a precise description of the corresponding file.
    \item double duration: array with the duration of each period.
    \item cv\_weights\_file: path of a file containing the weights to be used in cross validation. See below for a precise description of the corresponding file.
    \item relax\_empirical\_fix: flag that, when set, relaxes the constraint that makes the sum of estimated rates per class equal to the empirical rate by class.
\end{itemize}

Note that to use the model with covariates, the user only needs to specify the info, arrivals, neighbors, and duration files. Additionally, for the model without regressors, the user must specify the files alpha\_regions\_file and time\_groups\_file. When performing cross validation, the file cv\_weights\_file is also needed.

The file info\_file must contain two lines. The first line must contain, separated by spaces, the number of periods \(T\) per day, the number of week days \(G\), the number of zones \(R\), the number of arrival types \(C\), the number  of covariates \(J\) and the number of holidays \(H\) for each year. The second line contains, for each day of the week and holiday, the number of observations \(N_{g}\) for that day or holiday. Values of index \(g\) from 0 to 6 are used for days of the week while holidays have indexes in \(\{7,8,...,7+H\}\).

The file arrivals\_file contains the samples observed, where each line describes a sample. Each line contains, separated by spaces, the period \(t \in [0,T)\), the day \(g \in [0,G)\), the zone \(r \in [0,R)\), the arrival type \(c \in [0,C)\), the sample index \(j \in [0,N_{g})\), the number of arrivals observed, and a flag indicating whether the day is a holiday or not.

The file neighbors\_file contains the description of each zone. Each line contains the zone index, two values representing the coordinates of the zone centroid, a number indicating the zone's land type, J values corresponding to the values of all $J$ covariates, followed by a set of pairs of values
(index,distance)
indicating the distances of the considered zone to its neighbors (identified by their indexes).

The file alpha\_regions\_file describes the space penalization parameters. It contains a \(R \times R\) matrix (with entries separated by spaces) where entry \((i,j)\) corresponds to the weight in the objective function of the model without covariates if zones 
\(i\) and \(j\) are neighbors.

The file time\_groups\_file contains the description of time groups and their corresponding penalties. The first line must have the number \(\mathcal{G}\) of time groups. The subsequent \(D*T\) lines must have the group index of each period.  Then the subsequent \(\mathcal{G}\) lines may contain the weight of each group. If the method is set to "cross\_validation", the reading of time weights is skipped.

File cv\_weights\_file contains the weights to be used in cross validation, separated by spaces. Each weight must be non-negative.

The durations file contains the duration of each time index, in hours. The first line contains, separated by spaces, the number of day indexes and the number of time indexes.  

The LASPATED app is only supported on Linux operating systems. The user must have the GCC compiler and the Boost library. Libraries xtl and xtensor are also used and provided. In order to use the model with covariates, the user must have Gurobi installed. The directory Model\_Calibration/Cpp contains a Makefile that is used to compile the app with:

\begin{verbatim}
    $ make GUROBI_LIB_VER=-lgurobi110 GUROBI_HOME=/opt/gurobi1101/linux64 
    INCLUDE_BOOST=/usr/include/boost
\end{verbatim}

The Makefile uses GUROBI\_HOME to find Gurobi directory, GUROBI\_LIB\_VER to find the Gurobi C library version and INCLUDE\_BOOST to find the Boost directory. The values displayed above are actually the default values set in the Makefile, so if Boost and Gurobi are installed in those locations, then the make command will compile the code. Once laspated app is successfully compiled, it can be run with:

\begin{verbatim}
    $ ./laspated -f config
    $ ./laspated -f --model_type=no_reg --method=calibration 
        --info_file=info.txt --arrivals_file=arrivals.txt 
        --neighbors_file=neighbors.txt 
        --alpha_regions_file=alpha_regions.txt
        --time_groups_file=time_groups.txt 
        --durations_file=durations.txt
        --output_file=output.txt
\end{verbatim}

where config is the config text file with options. Note that, in the second example, the options provided in the command line override any options with the same name in the config file.

After running laspated app, the output is saved in the path provided by option output\_file. If the model without covariates is used, the output contains on each line the indexes \(c \in [0,C)\), \(r \in [0,R)\), \(t \in [0,D*T)\), and the intensity \(\lambda_{crt}\) estimated by LASPATED. If the model with covariates is used, then each line correspond to indexes \(c \in [0,C)\), \(d \in [0,D)\), \(t \in [0,T)\), \(j \in [0,J)\), and the parameter \(\beta_{cdtj}\) estimated by LASPATED.

\section{Demo scripts}

In the LASPATED source code repository, we provide a demo directory, containing examples for the discretization and calibration functions. The demo summarizes all code excerpts provided in the previous sections. The directory contains three Python files:

\begin{itemize}
    \item demo\_discretization.py: Contains examples of how to generate a discretization;
    \item demo\_calibration\_regularized.py: Contain an example of running the calibration and cross-validation using the regularized model (without covariates);
    \item demo\_calibration\_covariates.py: Contain an example of running the calibration model with covariates.
\end{itemize}

The discretization demo can be run with

\begin{verbatim}
    $ python demo_discretization.py border_type discretization_type
\end{verbatim}

where border\_type can be: \textit{rectangle}, \textit{convex} or \textit{custom} and discretization\_type can be: \textit{rectangle}, \textit{hexagon} or \textit{custom}. The border type defines the type of border: rectangular, convex or the border of Rio de Janeiro city (custom). The discretization type can be rectangular, hexagonal or according to Rio de Janeiro's administrative regions (custom). The script considers arrival data from Rio de Janeiro and produces output files considering three time discretizations, and the Rio de Janeiro population as an additional feature.

Before running the calibration demo, the user must compile the laspated executable, following instructions from section~\ref{sec:laspated_cpp}. The calibration demo for the model without covariates can be run with

\begin{verbatim}
    $ python demo_calibration_regularized.py
\end{verbatim}

Additionally, the calibration demo for the model with covariates can be run with:

\begin{verbatim}
    $ python demo_calibration_covariates.py
\end{verbatim}

The input files for the C++ functions and the output of the functions are saved inside the demo/calib\_data respective subdirectories. The regularized model script runs the calibration function considering an unitary weight time and space regularization, and a cross-validation procedure considering a set of weights.

\section{Replication script}

Along with the LASPATED source code, we also provide a python script that performs all experiments from~\cite{laspatedpaper}. Before running the script, the first step is to compile the C++ code. If you have a GCC compiler installed,  this can be done by running the command:
\begin{verbatim}
    $ cd LASPATED/Replication 
    $ make -C cpp_tests
\end{verbatim}

The cpp\_tests directory contains a Makefile that can be edited by the user that compiles the C++ script.  The Makefile inside directory cpp\_tests accesses the user-defined environment variable \$GUROBI\_HOME. If it is set, then the code for the model with Covariates inside laspated.h is accessible. Otherwise, the script will not run the experiments that use covariates.

After the C++ code is successfully compiled, you can run the replication script by simply running:

\begin{verbatim}
    $ python replication_script.py
\end{verbatim}

This command will generate the rectangular, hexagonal and district discretizations using the laspated python module, run the experiments with the C++ code and process the results, generating the figures and tables presented in the paper.

All results presented in the paper are saved in the directory replication\_results, with subdirectories plots and tables. If you want access to the raw data generated by C++, check the directory cpp\_tests/results, with subdirectories for each experiment made. Subdirectories ex1 and ex2 contain results for the first and second model without covariates (Examples 1 and 2). Subdirectory ex3 contains results for the model with covariates (Table 2 of the paper) and real\_data contains results for the experiments with real data of Rio de Janeiro. For more details, see~\cite{laspatedpaper}.

\section{Docker container} \label{sec:docker}

We also provide a Docker container. It is available at \url{https://dockerhub.com} or via the Dockerfile in the repository. The docker container comes with all dependencies installed and the library ready to use. 

\subsection{Dockerfile}

To build the container via the repository Dockerfile, run:

\begin{verbatim}
    docker build -t laspated .
\end{verbatim}

If you have a Gurobi Web License, you can build the container with Gurobi support by running:
\begin{verbatim}
    docker build --build-arg USE_GUROBI=1 -t laspated .
\end{verbatim}

The above command will build the container with Gurobi 11.0.1 installed.

\subsection{DockerHub}

To download the container from DockerHub, just run:

\begin{verbatim}
    docker pull victorvhrn/laspated
\end{verbatim}

Note that the DockerHub container is built without Gurobi support by default. To enable Gurobi support, the C++ code must be recompiled inside the container. Check the next section on how to run the container and section~\ref{sec:laspated_cpp} on how to recompile the C++ code.

\subsection{Running the container}

To run the container, use:

\begin{verbatim}
    docker run -it laspated
\end{verbatim}

To run the container with Gurobi support, you can pass a valid Gurobi Web license to the container with:
\begin{verbatim}
    docker run --volume="/absolute/path/to/gurobi.lic:/opt/gurobi/gurobi.lic:ro" 
    -it laspated-dock 
\end{verbatim}

Both commands will open a shell environment with all dependencies installed. Once in the container environment, you can run both the Python and C++ functions.

\section{Missing Data}

An example of model estimation with missing data is available in subdirectory Missing\_Data. The subdirectory contains the implementation of four models that consider missing location data, and one artificial example considering missing timestamp data.

The file missing\_data.hpp, implements the regularized model (\textbf{class MissingLambdaRegularizedModel}) and a model that assumes the probability of a missing location for an arrival is proportional to the population in each zone (class \textbf{MissingLambdaCovariatesModel}). Additionally, missing\_data.hpp implements a customized version of the cross validation procedure for the regularized model. 

The file missing\_data.cpp implements a function that performs the estimation using a closed form expression, a function that uses the Fisher information matrix to compute confidence intervals, a function that runs an artificial example with missing timestamps, and a function that runs a set of experiments on the models. This code corresponds to the tests performed in 
the numerical experiments section of \cite{ref_missing_data}.

Because of the Fisher information matrix computation, the missing data code requires the xtensor-blas and the openblas libraries. To compile the code, run:

\begin{verbatim}
    cd Missing_Data
    g++ -o missing Cpp/missing_data.cpp -DUSE_GUROBI=1 -std=c++14 -m64 
    -I../Model_Calibration/Cpp -I$GUROBI_HOME/include -L$GUROBI_HOME/lib 
    -lboost_program_options -lopenblas -O3 -lgurobi110 -lgurobi_c++
\end{verbatim}

The signature of the cross validation function for missing data is given by:

\begin{verbatim}
laspated::CrossValidationResult cross_validation(
    laspated::Param& param, MissingLambdaRegularizedModel& model,
    xt::xarray<int>& sample, xt::xarray<int>& sample_missing_arrivals,
    std::vector<double>& group_weights) 
\end{verbatim}

The following example shows how to run the C++ code for missing data:

\begin{verbatim}
    ./missing -f test.cfg --model=all --info_file=info.dat 
    --neighbors_file=neighbors.dat --arrivals_file=arrivals.dat 
    --missing_file=missing_arrivals.dat
\end{verbatim}

This example runs all missing data models for the input files provided. In general, options can be provided in a configuration file, the test.cfg file provided with -f, or in the command line. Note that the command line options override the values in the configuration file. The following options are identical to the LASPATED application: 
\begin{itemize}
    \item EPS,
    \item sigma,
    \item max\_iter,
    \item lower\_lambda,
    \item beta\_bar,
    \item info\_file,
    \item arrivals\_file,
    \item neighbors\_file.
\end{itemize}

The remaining options are specific to the missing data application. The option missing\_file specifies the path to the file containing missing arrivals' data. This file follows the same format as arrivals\_file, but the application ignores the value of the zone index. Option mc\_samples specifies the Monte Carlo sample size $S$, used in the \textbf{MissingLambdaCovariatesModel}. The option test\_weights contains a list of weights to be used in the \textbf{MissingLambdaRegularizedModel}.

The option model defines the missing data model(s) that the application will run. It can be one of the following: 
\begin{itemize}
    \item all: runs all four missing data models;
    \item analytical: runs only the closed form models;
    \item regularized: runs only the \textbf{MissingLambdaRegularizedModel}, once for each of the test\_weights set, and once performing a cross validation;
    \item population: runs only the \textbf{MissingLambdaCovariatesModel}, with the number of samples provided by mc\_samples.
\end{itemize}

For the \textbf{MissingLambdaCovariatesModel}, there is also the function mean\_var\_u, that computes the mean and variance of the Monte Carlo samples for an observation indexed by $c \in \mathcal{C}, d \in  \mathcal{D}, t \in \mathcal{T}, n \in \mathcal{N}_{c,d,t}$. Its signature is:

\begin{verbatim}
    std::pair<double, double> mean_var_u(xt::xarray<int>& mn_samples, int c,
                                       int d, int t, int n,
                                       xt::xarray<double>& x)
\end{verbatim}
where mn\_samples is the $S \times |\mathcal{I}|$ array of samples, $(c,d,t,n)$ is the observation index and $x = (x_{c,d,t,i}: c \in \mathcal{C}, d \in \mathcal{D}, t \in \mathcal{T}, i \in \mathcal{I})$ is the Poisson intensities array. The function returns a pair containing the mean and variance, respectively.

\bibliographystyle{plainnat}
\bibliography{Biblio}

\end{document}